\documentclass[pra,twocolumn,aps,superscriptaddress,showpacs]{revtex4-1}

\usepackage{amsmath}
\usepackage{amssymb}
\usepackage{lineno}
\usepackage{graphicx}
\usepackage{color}
\usepackage{bm}
\usepackage{times}
\usepackage{pdfpages}
\usepackage{hyperref}
\hypersetup{
  colorlinks=true,
  citecolor=blue,
  linkcolor=blue,
  urlcolor=blue}

\everymath{\displaystyle}
\begin{document}

\title{Thermal suppression of phase separation in condensate mixtures}

\author{Arko Roy}
\affiliation{Physical Research Laboratory,
             Navrangpura, Ahmedabad-380009, Gujarat,
             India}
\affiliation{Indian Institute of Technology,
             Gandhinagar, Ahmedabad-382424, Gujarat, India}
\author{D. Angom}
\affiliation{Physical Research Laboratory,
             Navrangpura, Ahmedabad-380009, Gujarat,
             India}

\begin{abstract}
We examine the role of thermal fluctuations in binary condensate mixtures
of dilute atomic gases. In particular, we use Hartree-Fock-Bogoliubov with
Popov approximation to probe the impact of non-condensate atoms to the 
phenomenon of phase-separation in two-component Bose-Einstein condensates. 
We demonstrate that, in comparison to $T=0$, there is a suppression in the
phase-separation of the binary condensates at $T\neq0$. This arises from the
interaction of the condensate atoms with the thermal cloud. We also show that,
when $T\neq0$ it is possible to distinguish the phase-separated case from 
miscible from the trends in the correlation function. However, this is not
the case at $T=0$.

\end{abstract}



\maketitle


Phase-separation in a two-component fluid is ubiquitous in nature, and the
transition from miscible to immiscible phase is a quintessential example of 
critical phenomena. One classic example is the temperature driven 
phase-separation in the cyclohexane-aniline mixture ~\cite{stanley_71}. 
It is then natural to ask what are the similarities and differences in 
binary mixtures of quantum fluids ? The recent experimental advances in binary 
Bose-Einstein condensates (BECs) of dilute atomic gases provide an ideal 
testbed to address such a question. 
In the case of binary mixtures of BECs or two-species BEC (TBEC) tuning the
interaction through Feshbach resonances~\cite{inouye_98,chin_10} can render it 
miscible or immiscible. Using improved experimental techniques, TBECs have 
been achieved in mixtures of two different alkali atoms 
~\cite{modugno_02,thalhammer_08,lercher_11,mccarron_11,pasquiou_13}, or two 
different isotopes ~\cite{inouye_98,papp_08,tojo_10} and atoms of the same 
element in different hyperfine states ~\cite{stamper_kurn_98,stenger_98,
myatt_97,sadler_06} over 
the last decade. The remarkable feature of phase-separation in 
TBECs  has been successfully observed in $^{85}$Rb-$^{87}$Rb 
~\cite{papp_08,tojo_10} and $^{87}$Rb-$^{133}$Cs \cite{mccarron_11}
condensate mixtures. 

The criterion for phase-separation, derived from Thomas-Fermi (TF) 
approximation at 
zero temperature \cite{pethick_08}, is that the intra-($U_{11},U_{22}$) and 
inter-species interaction ($U_{12}$) strengths, must satisfy the inequality 
$U_{12}^2>U_{11}U_{22}$. However, experiments are conducted at finite 
temperatures, and therefore, deviations from the criterion is to be expected.
Theoretical studies on effects of thermal cloud on 
phase-separation have been carried out for 
homogeneous binary Bose gases using Hartree-Fock
theory~\cite{schaeybroeck_13} and large-$N$ approximation~\cite{chien_12}.  
Phase-separation of trapped binary mixtures 
at finite temperature has also been examined using local-density
approximation~\cite{shi_2000}.
In this Letter we address this issue by using 
Hartree-Fock-Bogoliubov theory with Popov approximation (HFB-Popov)
~\cite{griffin_96} to account for the thermal fluctuations. It is a gapless
formalism satisfying Hugenholtz-Pines theorem~\cite{hugenholtz_59} and can be 
employed to compute the energy eigenspectra of the quasi-particle excitations 
of the condensates.

The method has been validated extensively in single species BEC, and we have
used it in our recent works to examine the effect of quantum fluctuations in 
TBECs \cite{roy_14a}. In the present work, we systematically study the role
of thermal fluctuations in the phenomenon of phase-separation in 
trapped TBECs. Our
studies reveal that at $T\neq0$, the constituent species in the TBEC undergo
phase-separation at a higher $U_{12}$ than the value predicted based on the
TF-approximation at $T=0$. Consistent with experimental observations of 
dual species condensate of $^{87}$Rb and $^{133}$Cs~\cite{mccarron_11}, our 
theoretical investigations show that even when the phase-separation
condition is met, there is a sizable overlap between the two species. We
attribute this to the presence of the thermal cloud, which have profound
affects on the miscibility-immiscibility transition. At $T=0$, the TBECs are
coherent throughout the spatial extent of the condensate, however, when 
$T\neq0$ coherence decays and is reflected in the correlation function. 
This implies that at $T=0$, the miscible or immiscible phases are
indistinguishable from the trends in the correlation function. But, 
for $T\neq0$ the miscible-immiscible transition and the associated
changes in the density profiles have a characteristic signature in the form 
of the correlation functions. There is a smooth cross-over between the 
correlations functions when the transition occurs. Interspecies Feshbach 
resonances of ultracold bosons have been experimentally demonstrated for 
Na-Rb~\cite{wang_13}, K-Rb ~\cite{simoni-03} and Cs-Rb~\cite{pilch_09} 
mixtures, but Bose condensed mixture of Na-Rb is yet to observed 
experimentally. The Cs-Rb condensate mixture is a stepping stone
towards the production of quantum gas of dipolar RbCs molecules, as
unlike the KRb molecule, the rovibrational ground state of RbCs molecule is 
stable against exchange of atoms. Considering this we focus our study on the
finite temperature effects in the Cs-Rb condensate mixture. Other than 
tuning the interspecies, it is also possible to steer the condensate mixture
through the miscible-immiscible transition using intraspecies Feshbach 
resonance. An example is the tuning of the intraspecies interaction of 
$^{85}$Rb in $^{85}$Rb-$^{87}$Rb~\cite{papp_08,tojo_10}, and for this system
too, we have examined the suppression of phase-separation at 
$T\neq0$~\cite{smho}. It must be emphasized that, as the 
background scattering length of  $^{85}$Rb is negative, it is 
possible to obtain $^{85}$Rb BEC \cite{cornish_00} only with the use of 
Feshbach resonance \cite{roberts_98}.


{\em Theory} ---
We consider a cigar shaped TBEC, where the frequencies of the harmonic
trapping potential satisfy the condition 
$\omega_{\perp}  \gg \omega_z$ with $\omega_x=\omega_y=\omega_{\perp}$. 
In this case, the radial excitation energies are large and assume 
the radial degrees of freedom are frozen for which
$\hbar\omega_{\perp}\gg \mu_k$. So, the dynamics and hence the 
excitations occur only along the axial direction, $z$-axis, of the trap. In 
the mean-field regime, using HFB-Popov approximation~\cite{roy_14,roy_14a}, a
pair of coupled generalized 1D Gross-Pitaevskii (GP) equations describe the 
dynamics and density distribution of the TBEC. The combined form of the 
equations is
\begin{eqnarray}
 \hat{h}_k\phi_k + U_{kk}\left[n_{ck}+2\tilde{n}_{k}\right]\phi_k
  +U_{12}n_{3-k}\phi_k=0,
\label{gpem}
\end{eqnarray}
where $\hat{h}_{k} = (-\hbar^{2}/2m_k)\partial ^2/\partial z^2 +
V_k(z)-\mu_k$ is the one-body part of the Hamiltonian, with $k=1,2$
as the species label. The strength of the coupling constants are given by
$U_{kk} = (a_{kk}\lambda)/m_{k}$ and $U_{12}=(a_{12}\lambda)/(2m_{12})$,
where, for cigar shaped traps $\lambda =\omega_{\perp}/\omega_z \gg 1$.
Without loss of generality, for stable configurations the intra-species 
scattering lengths $a_{kk}$, and the inter-species scattering length $a_{12}$ 
are considered as positive (repulsive). Under the HFB approximation, the Bose 
field operators are decomposed as $\hat{\Psi}_k = \phi_k + \tilde{\psi}_k$,
where the $\phi_k$s are the stationary solutions of Eq.~(\ref{gpem}) obtained 
by evolving the solution in imaginary time, with 
$n_{ck}(z)\equiv|\phi_k(z)|^2$.  The field operator $\tilde\psi_{k}(z)$ 
represents the fluctuation part of $\hat{\Psi}_k(z)$, it incorporates both 
the quantum and thermal fluctuations. The fluctuation operator, both quantum 
and thermal, are functions of the elementary excitations of the system, 
which solves the coupled Bogoliubov-de-Gennes equations  
\begin{subequations}
\begin{eqnarray}
 \hat{{\mathcal L}}_{1}u_{1j}-U_{11}\phi_{1}^{2}v_{1j}+U_{12}\phi_1 \left 
   (\phi_2^{*}u_{2j} -\phi_2v_{2j}\right )&=& E_{j}u_{1j},\;\;\;\;\;\;\\
    \hat{\underline{\mathcal L}}_{1}v_{1j}+U_{11}\phi_{1}^{*2}u_{1j}-U_{12}
    \phi_1^*\left (\phi_2v_{2j}-\phi_2^*u_{2j} \right ) 
     &=& E_{j}v_{1j},\;\;\;\;\;\;\\
    \hat{{\mathcal L}}_{2}u_{2j}-U_{22}\phi_{2}^{2}v_{2j}+U_{12}\phi_2\left 
    ( \phi_1^*u_{1j}-\phi_1v_{1j} \right ) &=& E_{j}u_{2j},\;\;\;\;\;\;\\
\hat{\underline{\mathcal L}}_{2}v_{2j}+U_{22}\phi_{2}^{*2}u_{2j}-U_{12} 
\phi_2^*\left ( \phi_1v_{1j}-\phi_1^*u_{1j}\right ) &=& 
E_{j}v_{2j},\;\;\;\;\;\;\;\;\;
\end{eqnarray}
\label{bdg2m}
\end{subequations}
where $\hat{{\mathcal L}}_{1}=
\big(\hat{h}_1+2U_{11}n_{1}+U_{12}n_{2})$, $\hat{{\mathcal L}}_{2}
=\big(\hat{h}_2+2U_{22}n_{2}+U_{12}n_{1}\big)$ and
$\hat{\underline{\cal L}}_k  = -\hat{\cal L}_k$. Here $u_{kj}$s and $v_{kj}$s 
are the Bogoliubov quasi-particle amplitudes corresponding to the $j$th energy 
eigenvalue. The quantities 
$\tilde{n}_k(z)\equiv\langle\tilde{\psi}_{k}^{\dagger}(z,t)
\tilde{\psi}_k(z,t)\rangle$, and $n_k(z) = n_{ck}(z)+ \tilde{n}_k(z)$
are defined as non-condensate, and total density, respectively.
To solve the above eigenvalue equations, the $u_{kj}$s and
$v_{kj}$s are decomposed into a linear combination of harmonic oscillator
eigenstates. The order parameters $\phi_k$s and the non-condensate 
densities $\tilde{n}_k$s are then the self-consistent solutions of the
coupled Eqns.~(\ref{gpem}) and (\ref{bdg2m}). The thermal components, in terms
of the quasi-particle amplitudes, is 
\begin{equation}
 \tilde{n}_k=\sum_{j}\{[|u_{kj}|^2+|v_{kj}|^2]N_{0}(E_j)+|v_{kj}|^2\},
  \label{n_k2}
\end{equation}
where, $N_0(E_j) = (e^{\beta E_j} - 1) ^{-1}$ with $\beta=1/(k_{\rm B}T) $ is 
the Bose factor of the $j$th quasi-particle mode at temperature $T$.
A more detailed description of the 
decomposition and derivation of the relevant equations are given 
else where ~\cite{smho}. 
In this Letter we examine the role of temperature in phase-separation of 
TBECs. For this, a measure of phase separation  is the overlap integral
\begin{eqnarray}
  \Lambda = \frac{\left[\int n_{1}(z)n_{2}(z) dz\right]^2}
        {\left[\int n_{1}^2(z)dz\right]\left[\int n_{2}^2(z)dz\right]}.
\end{eqnarray}
Miscible phase is when $\Lambda=1$ and signifies complete overlap of the two 
species, whereas the binary condensate is completely phase-separated when
$\Lambda=0$~\cite{jain_11}.

In terms of the Bose field operator $\hat{\Psi}_k$, the normalized first order 
or the off-diagonal correlation function, which is  also a measure of the
phase fluctuations, is 
\begin{eqnarray}
 g^{(1)}_k(z, z') =
 \frac{\langle\hat{\Psi}_k^\dagger(z)\hat{\Psi}_k(z')\rangle}
 {\sqrt{\langle\hat{\Psi}_k^\dagger(z)\hat{\Psi}_k(z)\rangle
 \langle\hat{\Psi}_k^\dagger(z')\hat{\Psi}_k(z')\rangle}}.
\end{eqnarray}
It can also be expressed in terms of off-diagonal condensate and 
non-condensate densities as 
\begin{eqnarray}
g^{(1)}_k(z, z') = \frac{n_{ck}(z,z')+\tilde{n}_k(z,z')}{\sqrt{n_{k}(z)
n_{k}(z')}},
\label{corr}
\end{eqnarray}
where,
\begin{eqnarray}
n_{ck} (z,z')     & = & \phi_k^*(z)\phi_k(z') \nonumber \\
\tilde{n}_k(z,z') & = & \sum_{j}\{[u_{kj}^*(z)u_{kj}(z') 
                        + v_{kj}^*(z)v_{kj}(z')]N_{0}(E_j) \nonumber \\
                  &   & + v_{kj}^*(z)v_{kj}(z')\}.  \nonumber 
\end{eqnarray}
At $T=0$, when the entire system is coherent and characterized by the
presence of condensate only, then $g^{(1)}_k=1$ within the extent of the
condensate, whether it is in the miscible or in the immiscible regime.
So, one cannot distinguish between the two phases from the nature of
the correlation functions of the individual species. However, at $T\neq0$, a 
clear signature of miscible-immiscible transition of the density profiles is 
reflected in the form of the correlation functions.
\begin{figure}[t]
 \includegraphics[width=8.0cm]{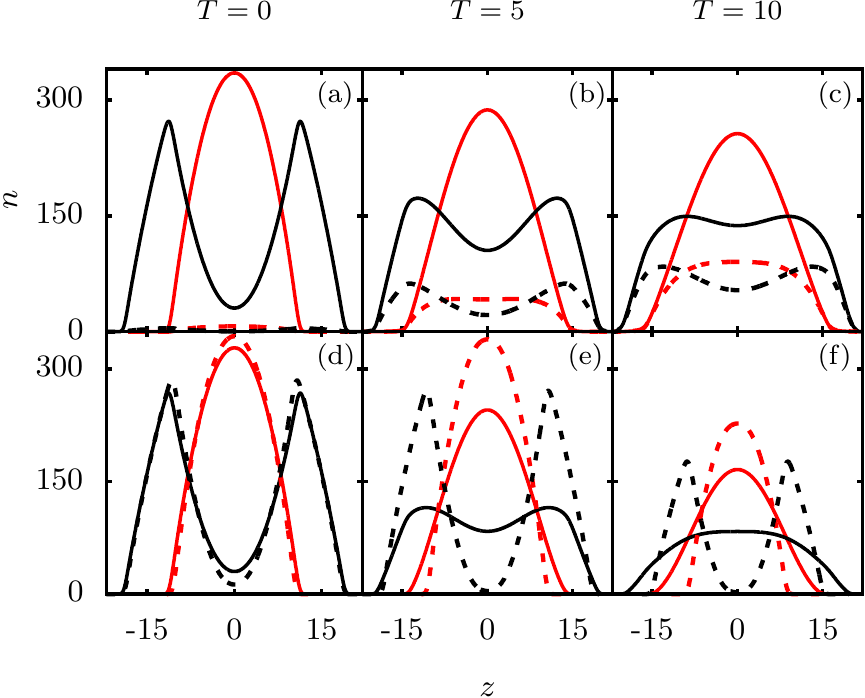}
    \caption{(Color online) 
             The suppression of phase-separation 
             in $^{87}$Rb-$^{133}$Cs
             TBEC at $a_{12}=295a_0$. (a)-(c) The solid and dashed 
             red (black) lines represent 
             $n_{\rm Cs} (n_{\rm Rb})$ and 
             $\tilde{n}_{\rm Cs} (\tilde{n}_{\rm Rb})$, respectively, at 
             $T = 0, 5, 10$nK. (d)-(f) The solid red (black) lines 
             represent $n_{c\rm Cs} (n_{c\rm Rb})$ at 
             $T = 0, 5, 10$nK respectively. The dashed red (black)
             lines $n_{c\rm Cs} (n_{c\rm Rb})$ at $T=0$ with the same 
             number of condensate atoms at $T = 0, 5, 10$nK respectively.
             Here, $n$ and $z$ are measured in units of $a_{\rm osc}^{-1}$ and
             $a_{\rm osc}$, respectively.}
    \label{profs_CsRb}
\end{figure}



  The thermal suppression of phase-separation is generic to any binary
condensate mixture. However, for comparison with experimental realizations
we consider the $^{133}$Cs-$^{87}$Rb BEC mixture with $^{133}$Cs labelled as 
species 1, and $^{87}$Rb as species 2. Here after, for brevity we drop the 
mass numbers and write these as Cs and Rb. The intra-species 
scattering lengths are $a_{11}=a_{\rm Cs}=280a_0$, $a_{22}=a_{\rm Rb}=100a_0$, 
the inter-species scattering length is $a_{12}=a_{\rm CsRb}=295a_0$ 
with $N_{\rm Cs} = N_{\rm Rb} = 5\times10^3$, and $a_0$ as the Bohr
radius. To form a quasi-1D trap we take 
$\omega_{z({\rm Cs})} = 2\pi\times 4.55 $Hz and 
$\omega_{z({\rm Rb})} = 2\pi\times 3.89 $Hz; 
$\omega_{\perp({\rm Cs})} = 50\omega_{z({\rm Cs})}$ and
$\omega_{\perp({\rm Rb})} = 50 \omega_{z({\rm Rb})} $. For this value of
$\omega_{\perp}$, the temperature along the radial direction is 
$\hbar\omega_{\perp}/k_{\rm B}\approx 11$nK, and the tight confinement
condition is valid as $\mu_k/\hbar\omega_{\perp} \approx10^{-2}$. In addition 
to this, the healing length $\xi_k\gg1/n_k$. Thus the system is in the weakly 
interacting TF regime \cite{petrov_00} and mean field description through 
GP-equation is valid. For this parameter set, the ground state density
distribution is phase-separated with species 1 at the center and surrounded 
by the species 2 at the edges. We refer to this configuration of density
profiles as {\em sandwich} type. This choice of parameters is
consistent with the experimental parameters of a recent work on quasi-1D TBEC 
of different hyperfine states of $^{87}$Rb \cite{de_14}, in which 
dynamical evolution of mixtures of quantum gases has been observed.
It should be emphasized here that {\em sandwich} type density
profiles are applicable only to trapped systems. In uniform systems, at
phase-separation, the energetically preferred states are the
symmetry-broken density profiles where one species is entirely to the left
and the other is entirely to the right. We refer to this configuration of
density profiles as {\em side-by-side} type. In the present work, we
demonstrate the role of thermal cloud in {\em sandwich} type density
profiles since these are unique to trapped systems and experimentally
pertinent. For the homogeneous binary condensates, using periodic boundary 
condition with $\omega_z=0$ in our computations, we do get side-by-side 
density profiles at phase separation. As an example, the density profiles
are shown in Ref. \cite{smho} and these are consistent with the results 
reported in previous works \cite{schaeybroeck_08}. 
In the computations the spatial and temporal variables are scaled as 
$ z/a_{\rm osc( Cs )}$ and 
$\omega_{z({\rm Cs })}t$ to render the equations dimensionless.

At $T=0$, in TBECs, as mentioned earlier, the criterion for phase separation 
is $U_{12}> \sqrt{U_{11}U_{22}}$. With the parameters of Cs-Rb TBEC, 
consider keeping $a_{\rm Cs}$ and $a_{\rm Rb}$ fixed, but varying 
$a_{12}=a_{\rm CsRb}$ through a magnetic Feshbach resonance \cite{pilch_09}.  
The condition for phase-separation, using TF-approximation, is then
$a_{12}>261a_0$. When $a_{12}=0$, the TBEC is non-interacting and the two
species are completely miscible, in which case $\Lambda=1$. 
On increasing $a_{12}$, the extent of overlap between the 
two-species decreases, and hence $\Lambda$ decreases. For instance, at 
$a_{12}=50a_0$, $\Lambda = 0.97$ and it decreases monotonically with
$\Lambda\rightarrow 0 $ at complete phase-separation. At $a_{12}=295a_0$, 
just at the onset of phase-separation, $\Lambda=0.14$. As shown in
Fig.~\ref{profs_CsRb}(a), the density profiles corresponding to
the two-species have interfacial overlap, and the interaction parameters 
satisfy the phase-separation condition. Furthermore, at phase separation, 
$n_{c{\rm Cs}}(0)$ is maximum, whereas $n_{c{\rm Rb}}(0)\approx0$ and the 
species do not have significant overlap. In other words, Cs at the center
of the trap is flanked by Rb at the edges and $\Lambda\approx 10^{-1}$. It is 
also to be mentioned here that for phase-separation, there is considerable
difference between the values of $a_{12}$ derived from TF approximation 
and the numerical solution of GP equation. This can be
attributed to large gradients in condensate densities, which are ignored in
the  TF-approximation.
\begin{figure}[h]
 \includegraphics[width=8.0cm]{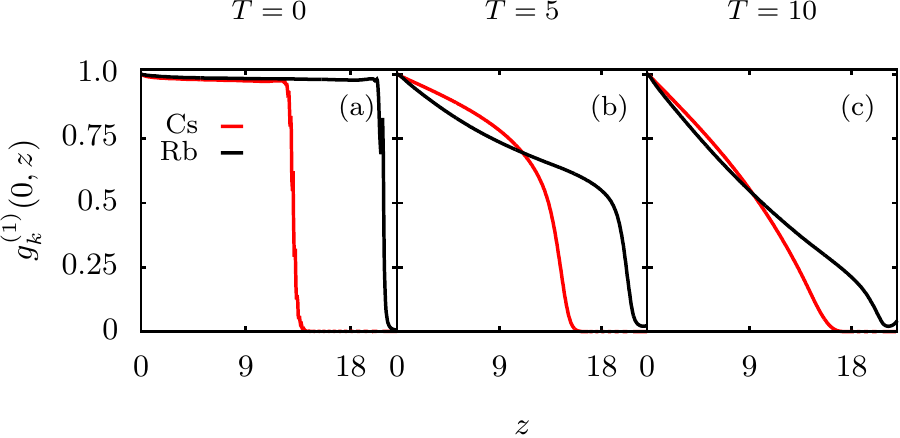}
    \caption{(Color online) (a)-(c) The first order spatial correlation 
             function, $g^{(1)}_{\rm Cs/Rb}(0,z)$ with $z \geqslant 0$, of 
             $^{87}$Rb-$^{133}$Cs TBEC at equilibrium 
             for $a_{12}=295a_0$ at $T = 0, 5, 10$nK respectively.
             Here $z$ is measured in units of $a_{\rm osc}$.}            
    \label{CsRb_corr}
\end{figure}


{\em Suppression of phase segregation} ---
For $T\neq0$ the Bose factor $N_0\neq0$, so in addition to the quantum 
fluctuations, the non-condensate densities $\tilde{n}_k$ have contributions 
from the thermal cloud as well. The condensate atoms $n_{ck}$ then interact 
with $\tilde{n}_k$ of both the species, and modify $n_{ck}$ of both the
species. For 
illustration, at $T=5$nK and $a_{12} = 295a_0$, the total and non-condensate 
density profiles are shown in Fig.~\ref{profs_CsRb}(b). Compared to the 
density profiles in Fig.~\ref{profs_CsRb}(a), there is a remarkable change
in $n_{c{\rm Rb}}$ as a result of the finite temperature: 
$n_{c{\rm Rb}}(0)>0$. Thus, keeping all the parameters same, but taking
$T=5$nK, the two species have substantial overlap as shown in
Fig.~\ref{profs_CsRb}(b) and $\Lambda$ becomes $\approx0.55$. In other words, 
the finite temperature transforms the phase separated TBEC at $T=0$ to a 
partially miscible phase. The degree of overlap increases with temperature
and at $T=10$nK, the TBEC is miscible as $\Lambda\approx0.77$. Thus, with 
the increase in temperature, the density of thermal cloud increases and 
the {\em phase-separation is suppressed}. This is evident from 
Fig.~\ref{profs_CsRb}(c), which shows the plots of corresponding total and 
non-condensate density profiles. Thus, $a_{12}$ has to be greater 
than $295a_0$ at $T\neq0$ for phase-separation to occur. To confirm that the
suppression is a consequence of non-zero temperature, we identify and compute 
the number of condensate atoms in each species, and use these numbers for 
$T=0$ computations. Despite the difference in the numbers of atoms,
from the plots in Figs.~\ref{profs_CsRb}(d-f), the TBEC retains the 
immiscible profiles at zero temperature. This implies that without the 
thermal cloud, there are no deviations from the usual phase-separation 
condition. For comparison, plots of $n_{ck}$ from the finite temperature 
cases are also shown in the figure.
\begin{figure}[h]
 \includegraphics[width=7.0cm]{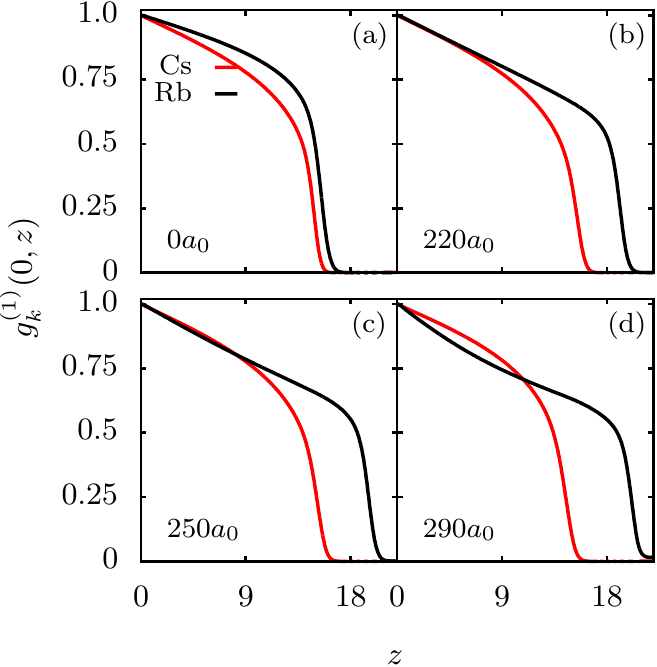}
    \caption{(Color online) (a)-(d) The first order spatial correlation 
             function, $g^{(1)}_{\rm Cs/Rb}(0,z)$ with $z \geqslant 0$, of 
             $^{87}$Rb-$^{133}$Cs TBEC at equilibrium
             at $T = 5$nK for $a_{12}=0, 220, 250, 290 a_0$,  respectively.
             Here $z$ is measured in units of $a_{\rm osc}$.}
    \label{combd_corr_CsRb}
\end{figure}

To investigate the spatial coherence at equilibrium, we examine the 
nature of the first order correlation function $g^{(1)}_k(z, z') $ as defined
in Eq. (\ref{corr}). As to be expected, profile of $g^{(1)}_k(z, z') $ 
depends on the interplay between the interaction strength and temperature. 
As stated earlier, at $T=0$, there is perfect coherence in the Cs-Rb TBEC and 
$g^{(1)}_{\rm Cs/Rb}(0,z)=1$ within the spatial extent of the condensates. 
This is independent of whether the TBEC is in miscible or immiscible regime. 
For simplicity and based on the symmetry of the system we consider 
$g^{(1)}_{\rm Cs/Rb}(0,z)$ with $z \geqslant 0$, and plots at different
temperatures are shown in Fig.~\ref{CsRb_corr}. At $T=0$ the form of the 
$g^{(1)}_{\rm Cs/Rb}(0,z)$ remains unchanged as the system undergoes the 
dramatic transition from miscible to immiscible phase. This is evident from 
the plot in Fig.~\ref{CsRb_corr}(a). However, when $T\neq0$, unlike the
zero temperature case, $g^{(1)}_{\rm Cs/Rb}(0,z)$ is maximum at $z=0$
and decays to zero with $z$. This is due to the non-condensate atoms, which
modify the the nature of coherence in the system. The rate of decay of the 
$g^{(1)}_{\rm Cs/Rb}(0,z)$ increases with temperature, and this is evident
from the plots of $g^{(1)}_{\rm Cs/Rb}(0,z)$ at $T=5$nK and $T=10$nK shown
in Figs.~\ref{CsRb_corr}(b-c) for $a_{12}=295a_0$. We also observe a dramatic 
variation in $g^{(1)}_{\rm Cs/Rb}(0,z)$ at fixed temperature,
but the value of $a_{12}$ is steered from miscible to immiscible regime. 
At the outset, when the TBEC is miscible at $a_{12}=0$,
 $g^{(1)}_{\rm Rb}(0,z)$ decays to $0$ at a larger distance than 
$g^{(1)}_{\rm Cs}(0,z)$ as shown in Fig.~\ref{combd_corr_CsRb}(a). This is 
because $n_{\rm Rb}$ has a larger spatial extent than $n_{\rm Cs}$. As 
$a_{12}$ is increased, the TBEC undergoes a phase-transition from miscible to 
sandwich type density profiles. Along with this, the distance at which 
$g^{(1)}_{\rm Rb}(0,z)$ falls off to zero increases with increase in $a_{12}$. 
On the contrary, the distance at which $g^{(1)}_{\rm Cs}(0,z)$ falls off to 
zero decreases with increase in $a_{12}$. This causes the 
$g^{(1)}_{k}(0,z)$ of the individual species to cross each other at a
certain $z_0$. At $z_0$, the two species have equal 
$g^{(1)}_{\rm Cs/Rb}(0,z_0)$ and this is a characteristic signature of 
immiscible phase. These features are shown in Figs.~\ref{combd_corr_CsRb}(b-d).
It deserves to be mentioned here that $z_0$ increases, and $g^{(1)}_{\rm
Cs/Rb}(0,z_0)$ decreases with increase in $a_{12}$.
In addition, there is a dramatic difference in the decay rates of 
$g^{(1)}_{\rm Cs/Rb}(0,z_0)$; it is much faster in Cs. 
This is attributed to the fact that both $n_{c{\rm Rb}}$ and
$\tilde{n}_{\rm Rb}$ increase along $z$ within the bulk of Cs-Rb TBEC. 
Where as in Cs, around the origin $n_{c{\rm Cs}}$ decreases but 
$\tilde{n}_{\rm Cs}$ increases. This trend is similar with a single-species  
Cs condensate. The presence of Rb does not affect the nature 
of $g^{(1)}_{\rm Cs}(0,z)$ in Cs-Rb TBEC. Around the point of
phase-separation $n_{c{\rm Rb}}(0)$ in a Cs-Rb TBEC is distinctly different 
from single species Rb condensate, so is the nature 
of $g^{(1)}_{\rm Rb}(0,z)$~\cite{smho}.
\begin{figure}[t]
 \includegraphics[width=8.0cm]{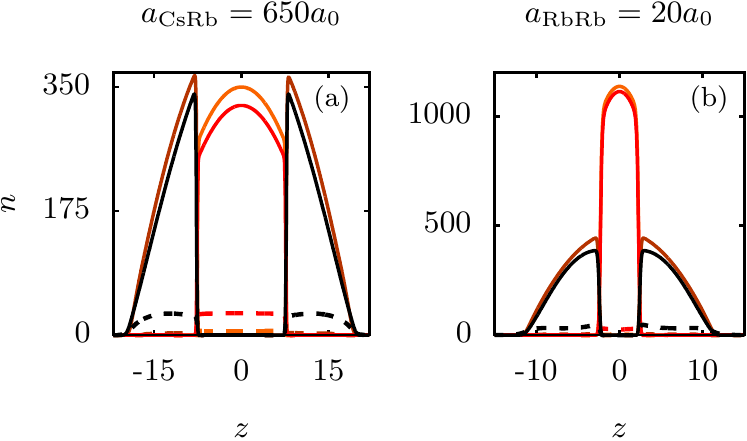}
    \caption{(Color online) 
             Density profiles showing complete phase-separation 
             at $T=0$ and $T=10$nK. (a) Phase-separation in 
             $^{87}$Rb-$^{133}$Cs TBEC for $a_{12}=650a_0$. The solid and
             dashed orange (brown) lines represent 
             $n_{c{\rm Cs}}(n_{c{\rm Rb}})$ and $\tilde{n}_{c{\rm
             Cs}}(\tilde{n}_{c{\rm Rb}})$, respectively at $T=0$. 
             The solid and dashed red (black) lines represent 
             $n_{c{\rm Cs}}(n_{c{\rm Rb}})$ and 
             $\tilde{n}_{c{\rm Cs}}(\tilde{n}_{c{\rm Rb}})$, respectively  
             at $T=10$nK. (b) Phase-separation in $^{85}$Rb-$^{87}$Rb TBEC
             for $a_{12}=20a_0$. The solid and dashed orange
             (brown) lines represent $n_c$ and $\tilde{n}$
             of $^{85}$Rb ($^{87}$Rb), respectively at
             $T=0$. The solid and dashed red (black) lines represent
             $n_c$ and $\tilde{n}$ of $^{85}$Rb ($^{87}$Rb), respectively 
             at $T=10$nK. Here, $n$ and $z$ are measured in units of 
             $a_{\rm osc}^{-1}$ and $a_{\rm osc}$, respectively.
            }
    \label{no_phase}
\end{figure}

{\em Segregation independent of temperature} ---
In the domain of large $a_{12}$, $U_{12} \gg \sqrt{U_{11}U_{22}}$,
 $\Lambda\approx0$, and the phase-segregation is more prominent. However, 
due to the geometry of the TBEC mean-field approximation is still valid. In this
domain the interfacial overlap is minimal and the TBECs assume sandwich type 
density profile. The system is then equivalent to three coupled condensate 
fragments, and as a result, the Bogoliubov analysis shows the presence of 
three Goldstone modes \cite{roy_14}. For the Cs-Rb TBEC considered here, the 
background inter-species scattering length $a_{\rm CsRb}=650a_0$ satisfies the 
above condition. With this value of $a_{12}$, at $T=0$ 
as shown in Fig.~\ref{no_phase}(a), Cs condensate lies at the center of the 
trap and Rb condensate at the edges. So, at the center $n_{\rm Rb}(0)=0$ and
$n_{\rm Cs}(0)$ is maximum. With the increase in $a_{12}$, there
is a decrease in the number of non-condensate atoms arising from quantum 
fluctuations. This is a manifestation of smaller overlap between the 
condensates at the interfaces. On the contrary,
for a single species BEC, with the increase in intra-species interaction
strength, the number of non-condensate atoms due to quantum fluctuations
increases\cite{roy_14a}. When $T\neq0$, the thermal density $\tilde{n}_{k}$ 
interact with the condensate clouds through the intra- and inter-species 
interactions. But, due to the large $a_{12}$, the inter-species 
interaction energy is much larger than the intra-species interaction energy. 
This makes $n_{\rm Rb}(0)\approx0$, and there is little overlap of the 
thermal cloud of one species with the condensate of the other species,
such that $\Lambda<0.1$. Thus, there is {\em no thermal suppression} in the
the domain of large $a_{12}$. We observe similar results in the 
case of $^{85}$Rb-$^{87}$Rb TBEC as well, where the the intra-species
interaction of $^{85}$Rb is decreased to obtain completely 
phase-separated density profiles. These are shown in Fig.~\ref{no_phase}(b).


{\em Conclusions} --- At finite temperatures, to examine the 
properties of binary condensates in the neighbourhood of phase separation, it 
is essential to incorporate the thermal component. In general, there is a 
delay or suppression of phase-separation due to the thermal component, and we 
have examined this in detail with the Cs-Rb binary condensate as an example. 
In this system the transition is driven by tuning the inter-species 
interaction, and similar results are obtained in $^{85}$Rb-$^{87}$Rb binary 
condensate, where tuning the intra-species interaction of $^{85}$Rb induce
the transition. The binary condensate mixtures of dilute atomic gases are
different from the classical binary fluids which undergo miscible-immiscible
transition with temperature as control parameter. First, the variation of 
temperature in TBECs is applicable only below the lower of the two critical 
temperatures.  Second, each species has two sub-components, the condensate
and non-condensate atoms. The condensate or the superfluid components are 
coherent, but the non-condensate components 
are incoherent and like the normal gas. Third, there are spatial density 
variations of all the components due to the nature of the confining potential 
and diluteness of the atomic gas. Fourth, beyond a certain critical value of
interaction strength or in the $U_{12} \gg \sqrt{U_{11}U_{22}}$
domain, temperature does not alter the density profiles. Finally, the 
transition to the phase separated domain at finite temperatures is 
associated with a distinct change in the profile of the correlation function.
Our results provide an explanation of the experimentally
observed density profiles at phase-separation\cite{mccarron_11}. Even when
the phase-separation condition is satisfied there is a finite overlap
between the two species at finite temperature which is due to the presence
of thermal cloud. Our findings clearly demonstrate the dominance of thermal 
fluctuations over quantum fluctuations which causes the suppression of
phase-segregation in any TBEC experiment.


\begin{acknowledgments}
We thank K. Suthar, S. Chattopadhyay and S. Gautam for useful
discussions. The results presented in the paper are based on the
computations using the 3TFLOP HPC Cluster at Physical Research Laboratory,
Ahmedabad, India.
\end{acknowledgments}

\bibliography{bec}{}
\bibliographystyle{apsrev4-1}

\end{document}